\documentclass[aps,pra,amsmath,amssymb,twocolumn,superscriptaddress,notitlepage]{revtex4-2}
\usepackage{graphicx}
\usepackage{color}
\usepackage{braket}
\definecolor{burntorange}{rgb}{0.8, 0.33, 0.0}\usepackage[colorlinks,citecolor=blue,linkcolor=red,urlcolor=blue]{hyperref}
\hypersetup{
  pdfkeywords={},
  pdftitle={Optimal quantum trajectories for sensing},
  }

\makeatletter

\newsavebox{\@brx}
\newcommand{\llangle}[1][]{\savebox{\@brx}{\(\m@th{#1\langle}\)}%
  \mathopen{\copy\@brx\kern-0.5\wd\@brx\usebox{\@brx}}}
\newcommand{\rrangle}[1][]{\savebox{\@brx}{\(\m@th{#1\rangle}\)}%
  \mathclose{\copy\@brx\kern-0.5\wd\@brx\usebox{\@brx}}}
\makeatother

\begin{document}

\title{Biasing quantum trajectories for enhanced sensing}

\author{Theodoros Ilias}
\email{ti9792@princeton.edu}

\affiliation{
Department of Electrical and Computer Engineering, Princeton University, Princeton, New Jersey 08544, USA}


\begin{abstract}
Quantum continuous measurement strategies consist an essential element in many modern sensing technologies leading to potentially enhanced estimation of unknown physical parameters. In such schemes, continuous monitoring of the quantum system which encodes the parameters of interest, gives rise to different quantum trajectories conditioned on the measurement outcomes which carry information on the parameters themselves. Importantly, different trajectories carry different amount of information i.e they are more or less sensitive to the unknown parameters of interest. In this work, we propose a novel approach on how to systematically engineer the quantum open-system dynamics in order to increase the probability of obtaining trajectories of high sensitivity. We focus on the simplest case scenario of a single two level system interacting with ancillas which are in turn measured consisting the discretized version of continuous monitoring. We analyze the performance of our protocol and demonstrate that it may lead to a substantial enhancement of sensitivity, as quantified by the classical Fisher information, even when applied to such small system sizes, holding the promise of direct implementation to state-of-the-art experimental platforms and to large, complex many-body systems.
\end{abstract}
\maketitle

\section{Introduction}
Quantum metrology consists one of the most advanced quantum technologies to date; it holds the promise of harnessing  properties of quantum systems for enhanced parameter estimation~\cite{Giovannetti1330,PhysRevLett.96.010401} with practical advancements in many scientific and technological fields such as biology~\cite{TAYLOR20161,wu2016diamond}, medicine\cite{Rej2015,blinder202413}, chemistry~\cite{s90907266}, physics~\cite{PhysRevA.46.R6797,Moser2013,RevModPhys.87.637,Ilias2024}, and navigation~\cite{PhysRevApplied.12.014019}.  This has served as a motivation for the development of a plethora of different sensing protocols that range from entanglement-based quantum metrology~\cite{PhysRevD.23.1693,PhysRevA.47.5138,PhysRevA.50.67,PhysRevLett.79.3865}, including various variational approaches~\cite{Koczor_2020,PhysRevX.11.041045,Marciniak2022}, to criticality-enhanced sensing~\cite{PhysRevA.78.042105,PhysRevA.88.021801,PhysRevX.8.021022,PhysRevA.96.013817,PhysRevLett.123.173601,PhysRevLett.124.120504,dicandia2021critical,PhysRevLett.126.010502,Salado_Mej_a_2021,PRXQuantum.3.010354} and to quantum estimation via sequential measurements~\cite{PhysRevLett.96.010401,Burgarth_2015,PhysRevLett.117.160801,Sekatski_2017,PhysRevLett.113.250801} among others.

\begin{figure}[h!]
\centering{} \includegraphics[width=0.48\textwidth]{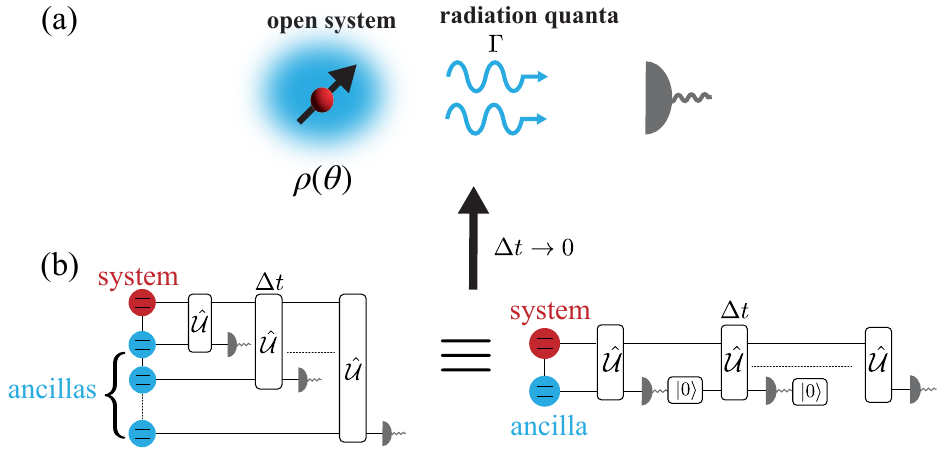} 
\caption{Collisional model approach to continuous measurement schemes. (a) Continuous measurement of the radiation quanta emitted by an open quantum system interacting with its bath. Each continuous fluorescent signal corresponds to a quantum trajectory which carries information on the unknown parameter, $\theta$, encoded on the state of the open-system acting as a sensor. An ensemble average over all such trajectories leads to an evolution of the sensor described by a standard Lindblad master equation (see main text). (b) Collisional model describing the discrete Markovian time dynamics of a quantum system. Left panel: At each instance the system of interest (in red), here a single qubit,  ``collides" with a different ancillary two-level system  (in blue), always prepared in the same ground state, for a time interval $\Delta t$. In turn, the corresponding ancilla is measured in the computational basis giving rise to a conditional discretized stochastic evolution of the system. Right panel: Instead of multiple ancillae, an equivalent description involves a single ancilla that is repeatedly re-initialized to the ground state after each measurement. In both cases, when the system-ancilla interaction is weak and occurs over infinitesimal time intervals (see main text), the discrete-time dynamics converges to a continuous-time limit, recovering the behavior described in panel (a).} 
\label{fig:fig1} 
\end{figure}

In the latter, the quantum system of interest interacts for some time interval, $\Delta t$, with some ancillary systems which are in turn measured; such process is repeated sequentially with the important detail that the system is \textit{not re-initialized}  to the same input state. In the limit where the system--ancilla interaction is short and weak, it leads to a continuous-type evolution of the system alone described by continuous master equations (MEs) as analyzed by collisional models~\cite{CICCARELLO20221} and illustrated in Fig. \ref{fig:fig1}. Recently there is an increasing effort of developing sensing protocols based on such \textit{continuous measurement schemes}~\cite{PhysRevLett.112.170401,PhysRevA.87.032115,PhysRevA.87.032115,PhysRevA.89.052110,PhysRevA.94.032103,Catana_2015,PRXQuantum.3.010354,PhysRevX.13.031012} which may or have already found applications in gravitational wave detectors~\cite{PhysRevLett.116.061102,Danilishin2012} and optomechanical force sensors~\cite{RevModPhys.82.1155,RevModPhys.86.1391} among others. In such schemes the measurement of the ancillary systems generate the so-called quantum trajectories\cite{RevModPhys.70.101, 1998aibp, carmichael1993open, wiseman_milburn_2009,gardiner2004quantum,gardiner2} which importantly carry information on the dynamic evolution and thus the unknown parameters of the system to be estimated.

\begin{figure}[t!]
\centering{} \includegraphics[width=0.48\textwidth]{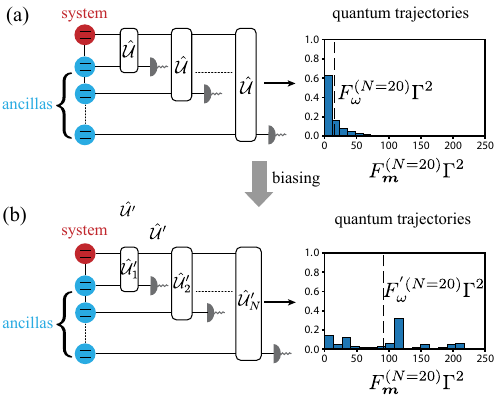} 
\caption{Biasing the quantum trajectories in the collision model approach. (a) The system of interest, a qubit in our toy model, collides sequentially with $N$ ancilla-qubits as described by the unitary interaction $\cal{U}$.  All ancillas are prepared in the same initial state and after each collision the corresponding ancilla is measured in the computational basis giving rise to a set of trajectories each characterized by sensitivity $F_{\boldsymbol{m}}^{(N)}$ (see main text) where we assume $N=20$.  In the right panel, it is shown the histogram of probability of finding $F_{\boldsymbol{m}}^{(N=20)}$.  With the dashed line it is shown the sensitivity of the system as quantified by the Fisher information (FI), $F_{\omega}^{(N)}$. (b) By properly engineering the system-ancilla interaction we can clearly bias the measurement outcomes towards quantum trajectories with higher sensitivity as shown in the right panel and resulting into higher FI and thus enhanced estimation capabilities. }
\label{fig:fig2} 
\end{figure}

 In this article, we provide a systematic way to engineer the system-ancilla interaction such that trajectories with higher sensitivity on small variations of the the unknown parameters of interest are more likely to occur, resulting eventually to improved estimation. Inspired by Refs.~\cite{PhysRevA.98.010103,Cilluffo_2021,PhysRevLett.131.120401} where the authors demonstrate how to alter the probability distribution of trajectories with certain properties, such as the measurement outcome of the ancillas and their temporal correlations, our protocol relies on assigning a corresponding sensitivity or precision to each of the measurement trajectories and biasing in favour of higher sensitivity as quantified by the classical Fisher information (FI). Using a simplified toy-model (see Fig. \ref{fig:fig2}) which can be straightforwardly implemented in various state-of-the-art quantum hardware platforms~\cite{Schindler2013, Arute2019, Evered2023,aquila}, our study paves the way for engineering nonequilibrium quantum states for enhanced sensing and it provides tools to control and study the effect of time correlations and memory effects on the metrological capabilities of complex quantum many-body systems.

\section{ Continuous measurement and collision models }

In this work, we are ultimately interested in estimating an unknown parameter, $\theta$, which is encoded in the dynamics of an open quantum system via continuous monitoring of the emitted radiation on the bath as illustrated in Fig.\ref{fig:fig1}(a). To be more specific and without the loss of generality, let us focus on the case where the parameter $\theta$ is encoded in the hamiltonian $\hat{H}_s(\theta)$ of the sensor, while the bath is monitored via detecting the fluorescence signal with detector efficiency $\eta=1$.  Under the typical assumptions underlying a quantum-optical master equation where the open-system is weakly coupled to a Markovian environment at zero temperature, the un-normalized conditional state of the sensor can be described in terms of a stochastic Schr{\"o}dinger equation \cite{RevModPhys.70.101,1998aibp,carmichael1993open,wiseman_milburn_2009,gardiner2004quantum,gardiner2}   
\begin{align} \label{eq:stochastic}
d \ket{\tilde{\psi}_c}=&-(i \hat{H}_s(\theta)+\sum_i \frac{\Gamma_i}{2} \hat{c}_i^{\dagger} \hat{c}_i) dt \ket{\tilde{\psi}_c}  \\
&+ \sum_i dN_i(t) (\sqrt{\Gamma_i dt}\hat{c}_i-\hat{1}) \ket{\tilde{\psi}_c}. \nonumber
\end{align}
Here $d N_i(t)$ is a stochastic Poisson increment accounting for the backaction of monitoring the emitted radiation quanta from the jump operator $\hat{c}_i$, and can take two values: $d N_i(t)=0$ with probability $p_0^{(i)}$ and $d N_i(t)=1$ with  with probability $p_1^{(i)}$. Particularly, in any infinitesimal time interval, $dt$, there is probability $p_1^{(i)}=dt \Gamma_i \langle \tilde{\psi}_c  |\hat{c}_i^{\dagger} \hat{c}_i | \tilde{\psi}_c \rangle/ \langle \tilde{\psi}_c \ket{\tilde{\psi}_c}$ of detected the radiation quanta associated with the jump operator $\hat{c}_i$ leading to the collapse of the conditional (unnormalized) state of the system $\ket{\tilde{\psi}_c} \rightarrow \sqrt{\Gamma_i dt} \hat{c}_i \ket{\tilde{\psi}_c}$. On the other hand, if there is no quanta detected with probability $p_0=\sum_i p_0^{(i)}=1-\sum_i p_1^{(i)}$ the system evolves according to the non-unitary evolution $\ket{\tilde{\psi}_c} \rightarrow \left( \hat{1}-dt (i \hat{H}_s(\theta) + \sum_i \Gamma_i \hat{c}_i^{\dagger} \hat{c}_i) \right) \ket{\tilde{\psi}_c}$. Repeating such a stochastic evolution defines a quantum trajectory, $D(t,0)$, consisting of the continuous fluorescent signal. An ensemble average over all conditional states leads to the definition of the density matrix $\rho_s(t) =\sum_{D(t,0)} \ket{\tilde{\psi}_c} \bra{\tilde{\psi}_c}$ whose evolution is governed by the standard Lindblad master equation (LME) \cite{breuer2002theory,carmichael1993open,wiseman_milburn_2009,gardiner2004quantum,Rivas_2012} 
\begin{equation}
\frac{d \rho_s}{dt}= -i [\hat{H}_s (\theta), \rho_s]+\sum_i \Gamma_i ( \hat{c}_i \rho_s \hat{c}_i^{\dagger}-\frac{1}{2} \{ \hat{c}_i^{\dagger} \hat{c}_i, \rho_s\} ).
\label{eq:LME}
\end{equation}
For the rest of this work, let us focus on the simplest case scenario where the system of interest is a single two-level system described by the Hamiltonian $\hat{H}_s= \omega \hat{\sigma}_x$ with $\omega \equiv \theta$ being the unknown parameter to be estimated and there is a single jump operator $\hat{c}_i \rightarrow \hat{\sigma}_-$ to be monitored. Here, $\hat{\sigma}_x$ is the Pauli-$x$ operator of the qubit and $\hat{\sigma}_{+ (-)}$ its raising (lowering) operator.

Motivated by the need to extend beyond the continuous-time processes discussed above, in this work we follow a collision model approach for the discrete Markovian time dynamics of the system as depicted in Fig.\ref{fig:fig1}(b). This framework not only provides the simplest but also the most natural microscopic description of continuous monitoring in open quantum systems. To be more specific,  in such model, the system, described in state $\ket{\psi}$, interacts or ``collides" with an ancillary two-level system according to the unitary evolution $\mathcal{\hat{U}}=e^{-i \hat{H} \Delta t}$, where $\hat{H}$ denotes the interaction Hamiltonian and $\Delta t$ the time duration of this interaction. In turn, the corresponding ancilla is measured in the computational basis, $\{ \ket{0}, \ket{1} \}$,  resulting in a stochastic evolution of the system described by the so-called Kraus operators $K_m=\bra{m}\mathcal{\hat{U}} \ket{0}$ where each ancilla has been initially prepared in the $\ket{0}$ state. Equivalently, such dynamics can be described by the use of a single ancillary system which is re-initilized after each collision with the system to its ground state~\cite{CICCARELLO20221}; cf. Fig \ref{fig:fig1}(b). In analogy to the continuous measurement scheme, the probability for the measurement outcome $m$ at each collision is given by the Born's rule $P_m=|| K_m \ket{\psi} ||^2$ and the system's evolution is updated according to $\ket{\psi} \rightarrow (K_m \ket{\psi})/\sqrt{P_m}$. We further note that in such description the ancillas are assumed to be initially uncorrelated, do not interact with each other and collide with the system only once; necessary conditions to guarantee a Markovian evolution of the system~\cite{CICCARELLO20221}. After $N$ collisions the quantum measurement trajectory $\boldsymbol{m}=(m_N,m_{N-1} \cdots m_1)$ is obtained with probability given by $P_{\boldsymbol{m}}=||K_{m_N}\cdots K_{m_1} \ket{\psi_0} ||^2$, where $\ket{\psi_0}$ is the initial state of the system. Finally, between two subsequent collisions the unconditional discrete time evolution of the system is described by the Kraus map $\rho_{N+1}=\mathcal{E}[\rho_N]=\sum_{m=0}^1 K_m \rho_N K_m^{\dagger}$ with $\sum_{m=0}^1 K_m^{\dagger}K_m= \hat{1}$. Under such framework, the dynamics of the qubit-sensor is described via interaction with the ancillas by the linear coupling interaction
\begin{equation}
\hat{H}=\omega (\hat{\sigma}_x \otimes \hat{1} )+\sqrt{\frac{\Gamma}{\Delta t}} (\hat{\sigma}_+ \otimes \hat{\sigma}_-+\hat{\sigma}_- \otimes \hat{\sigma}_+),
\label{eq:sys_anc_inter}
\end{equation}
Although henceforth we set $\Gamma=1/ \Delta t=1$ and $\omega=10 \Gamma$, we note that in the limit of $\Delta t \rightarrow 0$ and keeping terms up to first order in $\Delta t$, the evolution of the system's unconditional state follows the LME, Eq. (\ref{eq:LME}), describing the dynamics of a dissipative-driving qubit of strength $\omega$ and dissipation rate $\Gamma$. Meanwhile, the evolution of the conditional state is governed by the stochastic Schr{\"o}dinger equation in Eq. (\ref{eq:stochastic})~\cite{CICCARELLO20221}.

\section{Biasing trajectories for sensing}
\begin{figure}[t!]
\centering{} \includegraphics[width=0.48\textwidth]{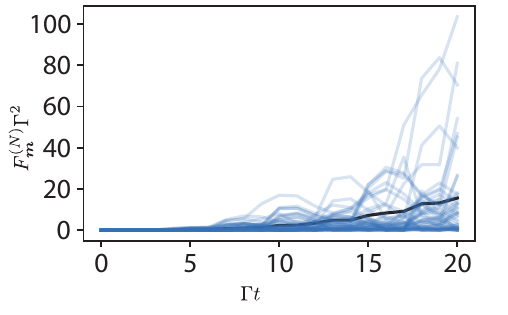} 
\caption{Precision of some of the individual trajectories (shown in blue transparent curves) as defined by Eq. (\ref{eq:FI_tr}) for our model described in the main text and for zero bias ($s=0$). With the black line it is shown the ensemble average over $10^4$ trajectories giving rise to the FI as in Eq. (\ref{eq:FI}).   }
\label{fig:fig3} 
\end{figure}

The starting point of our proposed sensing protocol lies in the definition of the Fisher information (FI):

\begin{equation}
F_{\omega}^{(N)}=\sum_{\boldsymbol{m}} P_{\omega}(\boldsymbol{m}) \{ \partial_{\omega} \log P_{\omega}(\boldsymbol{m})\}^2,
\label{eq:FI}
\end{equation}
as a figure of merit quantifying the metrological capabilities of our system since according to the Cram{\'e}r-Rao bound it sets a lower bound to the variance of any unbiased estimator ~\cite{Kay97}.  Here, $P_{\omega}(\boldsymbol{m}) \equiv P(\boldsymbol{m}/ \omega)$ denotes the probability of the measurement trajectory $\boldsymbol{m}$ conditioned on the unknown driving strength $\omega$ we aim to estimate with the sum is taken over all possible trajectories . As depicted in Fig.~\ref{fig:fig3}, the FI can be viewed as the ensemble average of sensitivity or precision of each individual trajectory i.e $F_{\omega}^{(N)}=\sum_{\boldsymbol{m}} P_{\omega}(\boldsymbol{m}) F_{\boldsymbol{m}}^{(N)}$ where we define:

\begin{equation}
F_{\boldsymbol{m}}^{(N)}= \{ \partial_{\omega} \log P(\boldsymbol{m}/ \omega)\}^2.
\label{eq:FI_tr}
\end{equation}
Note that in our model the measurement trajectory $\boldsymbol{m}=(m_N,m_{N-1} \cdots m_1)$ consists of a sequence of zeros and ones where each value indicates whether the corresponding ancilla was found in the ground or excited state, respectively.

Our goal is to engineer the system-ancilla interaction, $\mathcal{\hat{U}}_n \rightarrow \mathcal{\hat{U}}^{\prime}_n$, and steer the properties of the joint total system such that trajectories with high values of $F_{\boldsymbol{m}}^{(N)}$ are more likely to occur compare to the initial evolution as illustrated in  Fig.\ref{fig:fig2}(b). We note that while the initial interaction between the system and the ancilla is identical for each collision, $n$, i.e $\mathcal{\hat{U}}_n=\mathcal{\hat{U}}$ the engineered interaction may vary between collisions. Following Ref.~\cite{PhysRevLett.131.120401} and interpreting a measurement trajectory as a microstate of a fictitious spin system  such reweighing or biasing can be achieved by introducing a corresponding Boltzmann factor leading to the new deformed probabilities $P_{\omega}(\boldsymbol{m},s) \propto e^{-s \mathcal{B }} P_{\omega}(\boldsymbol{m})$ with $s$ being a ``global"  biasing factor assumed to be fixed and $\mathcal{B}$ a biasing function of our choice.  In the thermodynamic analogy with the spin system, $s$ plays the role of an inverse temperature while $\mathcal{B}$ corresponds to  an ``energy" function. While there is no restriction on the structure of $\mathcal{B}$, and we could, in principle, choose functions that incorporate bias based on time correlations of arbitrary order within the quantum trajectories, here we focus on the simplest case, where time correlations do not influence the reweighting. Consequently, we consider the form:

\begin{equation}
\mathcal{B}_{\boldsymbol{b}}(\boldsymbol{m})=\sum_{n=1}^N b_n m_n,
\label{eq:bias_function}
\end{equation}
with $\boldsymbol{b}$ being the different ``local" biasing factors to be determined. To generate the biasing ensemble $P_{\omega}(\boldsymbol{m},s)$, a tilted Kraus map can be used which for a single time step reads $\tilde{\mathcal{E}}_{n}=K_0 \rho_n K_0^{\dagger}+e^{-s b_n}K_1 \rho_n K_1^{\dagger} $. Although such map biases the system-ancilla dynamics in the desired way, it is not physically implementable since for $s b_n \neq 0$ it is not trace preserving. However, an alternative physical map can be constructed~\cite{PhysRevLett.131.120401} which depends on the collision number $n$  and consists of the Kraus operators:

\begin{equation}
\tilde{K}_0^{n}=G_n K_0 G_{n-1}^{-1},
\label{eq:tilted_Kraus0}
\end{equation}

\begin{equation}
\tilde{K}_1^{n}= e^{-s b_n/2}G_n K_1 G_{n-1}^{-1}.
\label{eq:tilted_Kraus1}
\end{equation}
Here, $G_{n-1}=\sqrt{\mathcal{E}_n^{*}[G_n^2]}$, where $G_N=\hat{1}$ and $\mathcal{E}_n^{*}[\bullet]=K_0^{\dagger} \bullet K_0+ e^{-s b_n} K_1^{\dagger}\bullet K_1$ is the dual tilted Kraus map. Finally, the initial state of the system needs to be modified according to $\ket{\psi_0} \rightarrow G_0 \ket{\psi_0} /|| G_0 \ket{\psi_0}||$ with $|| \bullet ||$ denoting the norm of the state. With this choice of the deformed Kraus operators the probability of a particular trajectory $\boldsymbol{m}=(m_N,m_{N-1} \cdots m_1)$ is given by $\tilde{P}_{\boldsymbol{m}}=|| \tilde{K}_{m_N}\cdots \tilde{K}_{m_1} \ket{\psi_0} ||^2/|| G_0 \ket{\psi_0}||^2=e^{-\sum_{i} s b_i m_i}|| K_{m_N}\cdots K_{m_1} \ket{\psi_0} ||^2/|| G_0 \ket{\psi_0}||^2$, thus introducing the desired bias in the relative probability ratio between two trajectories.  

Although the Stinespring dilation theorem \cite{stinespring,PRXQuantum.4.040309} guarantees that the new Kraus map defined by Eqs. (\ref{eq:tilted_Kraus0}) and  (\ref{eq:tilted_Kraus1}) can be unraveled into a new unitary collision model with an auxiliary ancilla, in practice for arbitrary initial system-ancilla interaction can be very challenging to implement. However, in Appendix \ref{app:small_collision_limit}, we show that in the infinitesimal limit of $\Delta t \rightarrow 0$ and under weak system-ancilla interaction, i.e in the continuous measurement limit, the new Kraus operators are $\tilde{K}_0^{n}=\hat{1}-i \omega \hat{\sigma}_x-\Gamma \Delta t/2 \hat{\sigma}_+ \hat{\sigma}_- e^{-s b_n}$ and $\tilde{K}_1^{n}=\sqrt{\Gamma \Delta t e^{-s b_n}} \hat{\sigma}_+ \hat{\sigma}_-$, which closely resemble the original map defined by ${K}_0=\hat{1}-i \omega \hat{\sigma}_x-\Gamma \Delta t/2 \hat{\sigma}_+ \hat{\sigma}_-$ and ${K}_1^{n}=\sqrt{\Gamma \Delta t } \hat{\sigma}_+ \hat{\sigma}_-$ respectively.  Consequently, the new map can be straightforwardly implemented using the same collisional model as the original, with only a time-dependent  modification of the system-ancilla interaction's strength, $\Gamma \rightarrow \Gamma e^{-s b_n}$. Importantly, such adjustment does not require any knowledge of the unknown parameter $\omega$. In the following, we describe two main strategies that can be followed for determining the biasing factors $b_n$ and thus the new tilted physical map resulting into enhanced precision.

\begin{figure}[h!]
\centering{} \includegraphics[width=0.48\textwidth]{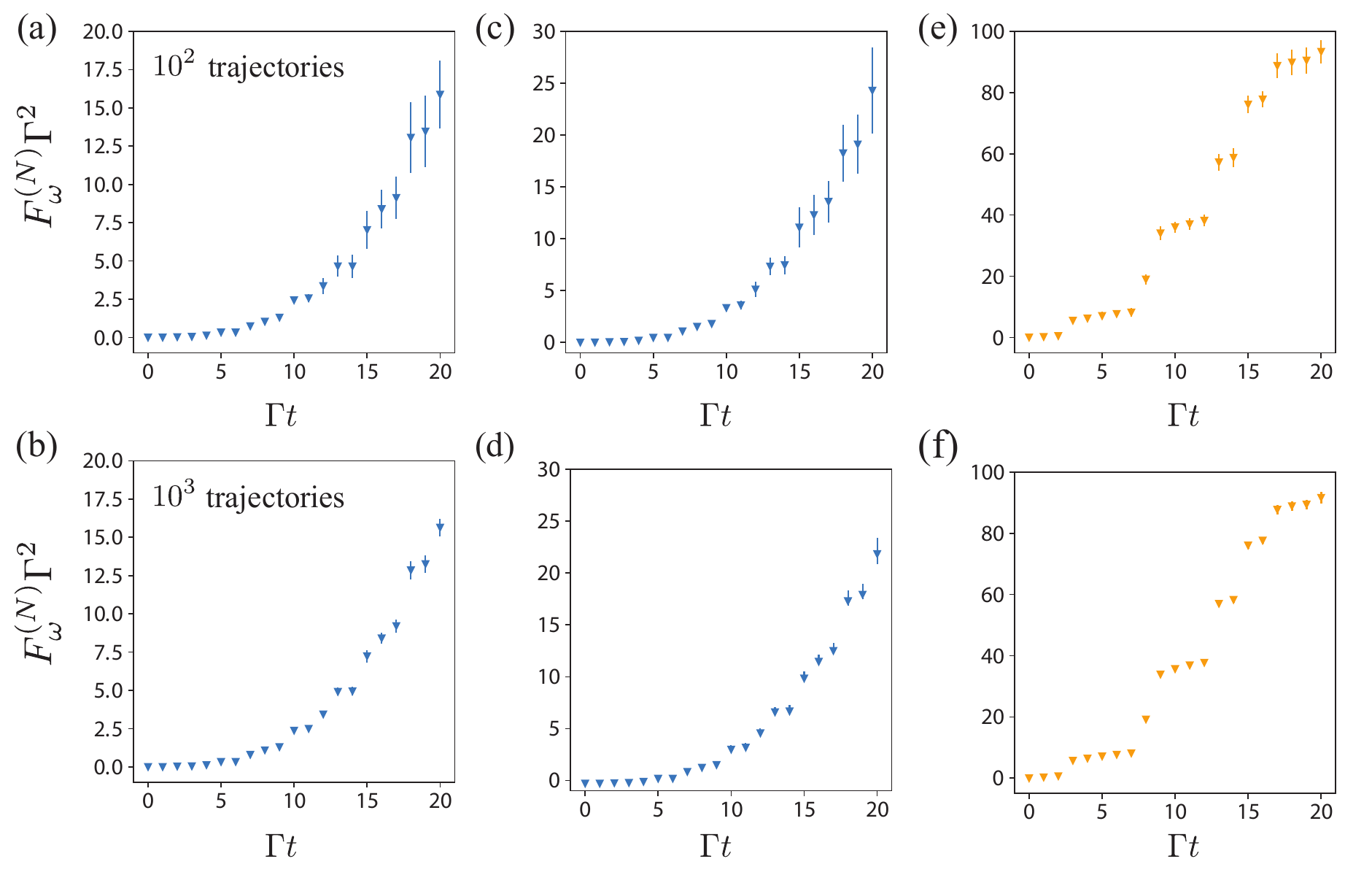} 
\caption{Convergence of the FI for different number of trajectories. In panels (a), (c) and (d) it is shown the numerically calculated FI for $10^2$ trajectories for the original Kraus operators, the ones obtained after the ``local" optimization algorithm for $s=1$ and for the ones obtained after the ``global" optimization algorithm for $s=3$ respectively. The error bars are estimated from the variance of ten independent samples, and are not shown if they are smaller than the data point size. Similar results are shown in panels (b), (d) and (f) for $10^3$ trajectories. }
\label{fig:fig4} 
\end{figure} 

\begin{figure}[t!]
\centering{} \includegraphics[width=0.48\textwidth]{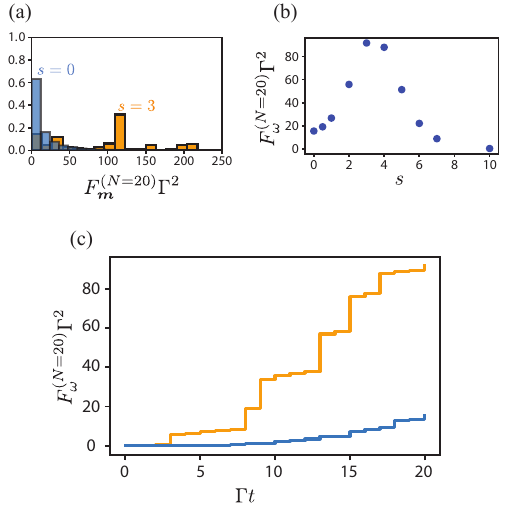} 
\caption{``Global" optimization approach. (a) Histogram of the precision of different quantum trajectories for the initial unbiased system-ancilla interaction i.e $s=0$ as shown in blue and for optimally biased trajectories according to the ``global" optimization protocol described in the main text ($s=3$) as illustrates in orange. The biasing increases the probability of trajectories with higher precision. (b) Dependence of the FI on the global optimization factor $s$. It is clear that there is an optimal value reflecting the fact that for $s$ very large there are few possible trajectories leading to low overall precision. (c) Evolution of the FI in time or in number of collisions for $s=0$ (in blue) and for the optimal value $s=3$ (in orange). In all panels the protocol is optimized for $N=20$ collisions and $10^4$ trajectories have been used for the calculation of FI.  }
\label{fig:fig5} 
\end{figure}

\emph{``Global" optimization} \textendash{}  As a first approach, we analyze a protocol that aims to reweigh the trajectories' ensemble leading to higher values of the FI after a specific, fixed number of collisions, $N$ which can be determined for example by the capabilities of the experimental platforms. More specifically,  the first step involves determining the quantum trajectory with the highest precision, $\boldsymbol{m}^{\rm max}$, by computing the FI and the precision of each trajectory in the initial map via the use of Eqs. (\ref{eq:FI}) and (\ref{eq:FI_tr}). Typically, such calculation can not be performed analytically and must instead be approximated by sampling a sufficient number of quantum trajectories, ensuring that the relative error remains within, $\epsilon_{\rm rel} \leq 1 \%$ after the $N$-th collision. The convergence of the FI for different number of trajectories is shown in Fig. \ref{fig:fig4}. After determining, $\boldsymbol{m}^{\rm max}$, of the original initial system-ancilla interaction, we choose $b_n=(-1)^{m_n}$ where $m_n$ is the outcome of measuring the ancilla after the $n$-th collision for the trajectory that achieves the highest precision. Consequently, our new constructed map is biasing the dynamics of our system towards trajectories that have as many similar outcomes as $\boldsymbol{m}^{\rm max}$. As shown in Fig.~\ref{fig:fig5}(c), even for our simple toy model, this results in a significant increase of the FI which however might not be optimized and in principle needs to be substantially modified when the number of collisions change even by a single instance. This is a consequence of the fact that the highest precision trajectory strongly depends on the total number of collisions and thus the given name ``global" optimization. The algorithmic steps that need to be followed in such approach are summarized below: 

\begin{center}
    \textit{``Global" optimization algorithm}
\end{center}

\begin{enumerate}
    \item Analytically calculate or numerically approximate by sampling sufficient number of quantum trajectories the FI in Eq.~(\ref{eq:FI}) up to the desired number of collisions. Compute the precision of each trajectory using Eq.~(\ref{eq:FI_tr}) and determine the one with the highest precision, $\boldsymbol{m}^{\rm max}$.
    \item Fix the values of the global and local biasing factors, $s$ and $b_n$ respectively depending on the desired strategy. As discussed above, although completely arbitrary in our scheme we choose $b_n=(-1)^{m_n}$ as the simplest possible case scenario.
    \item Construct the new physical map according to  Eqs.~(\ref{eq:tilted_Kraus0}) and (\ref{eq:tilted_Kraus1}).
    \item Similarly to step (1) compute, analytically or numerically, the FI of the biased dynamics.
\end{enumerate}

There are two elements that need to be emphasized in the procedure above. First of all, in the construction of the biased physical map the dependence on the unknown parameters lies not only on $K_0$ and $K_1$ but also on the hermitian matrices $G_n$. As a result, the precision of any particular single quantum trajectory before and after the reweighing has been actively modified. Furthermore, as depicted in Fig.~\ref{fig:fig5}(b) there is an optimal choice of how strongly we need to bias the ensemble of the trajectories for achieving the best precision. For $s\gg 1$, corresponding to strong bias, there are only few trajectories likely to occur and thus leading to very small overall precision. Clearly in the extreme case where there is only one probable trajectory, we obtain $F_{\omega} ^{(N)}=0$. It worths emphasizing that since each measurement trajectory is independent of the others, steps (1) and (4) of the ``global" optimization scheme can be massively parallelized, making the strategy feasible—albeit challenging—even for large system sizes or numerous ancillas corresponding to long evolution times.

\begin{figure}[t!]
\centering{} \includegraphics[width=0.48\textwidth]{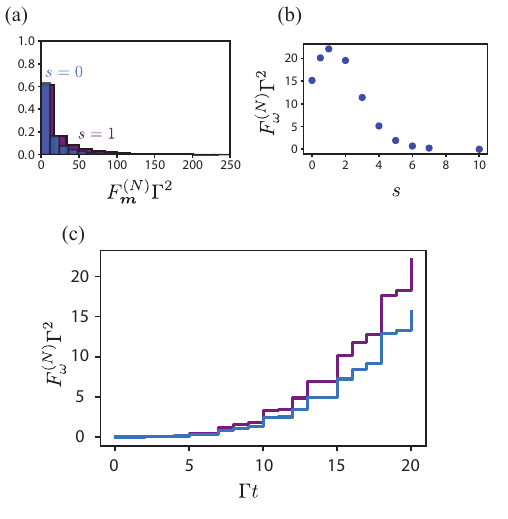} 
\caption{``Local" optimization approach. (a) Histogram of the precision of different quantum trajectories for the initial unbiased system-ancilla interaction i.e $s=0$ as shown in blue and for optimally biased trajectories according to the ``local" optimization protocol described in the main text ($s=1$) as illustrated in purple. Although the biasing increases the probability of trajectories with higher precision, the resulting histogram share many similarities to the one generated by the initial map (b) Dependence of the FI on the global optimization factor $s$. Similarly to the results of the ``global" optimization approach there is an optimal value of which importantly does not depend on the desired number of collisions we want to optimize for. (c) Evolution of the FI in time or in number of collisions for $s=0$ (in blue) and for the optimal value $s=1$ (in purple). $10^4$ trajectories have been used for the calculation of FI. }
\label{fig:fig6} 
\end{figure}

\emph{``Local" optimization} \textendash{} In contrast to the ``global" optimization approach described above and which in principle need to be repeated entirely each time the number of collision changes, here we present an alternative protocol that optimizes the FI and biases the quantum trajectories ensemble only up to the next collision i.e ``locally". Although such an approach consists a subset of the ``global" optimization algorithm and leads to a lower enhancement of the FI as demonstrated in Fig.~\ref{fig:fig6}, the corresponding tilted map retains much greater similarity to the original map and remains unchanged regardless of the number of collisions for which the precision of the scheme is being optimized. As a result, it requires less computational resources and thus it represents a promising alternative strategy for practical application of biasing the trajectories for  complex many-body systems. To be more specific, in such single-step bias approach for each collision we have $G_1=\hat{1}$ and $G_0^2=K_0^{\dagger} K_0+ e^{-s b_n} K_1^{\dagger} K_1$. Consequently, we can choose $b_n=-1$ and $b_n=1$ with initial state the post-collision state from the previous step and calculate the sensitivity of the measurement trajectory which corresponds to biasing towards the outcome where the ancilla is found in the excited state or the ground state respectively. At each step, we choose $b_n$ the one leading to the highest sensitivity. Repeating such process for $N$ desired collisions determines the target optimal trajectory $\boldsymbol{m}^{\rm max}$. Finally, after obtaining the optimal trajectory the associated new physical map can be constructed and the FI can be computed similarly to the ``global" optimization strategy described above. The steps of the ``local" optimization approach are the following:

\newpage

\begin{center}
    \textit{``Local" optimization algorithm}
\end{center}

\begin{enumerate}
    \item  Before each collision $n$ and given a global bias factor, $s$, construct the physical map according to Eqs.~(\ref{eq:tilted_Kraus0}) and (\ref{eq:tilted_Kraus1})  for the case $b_n=-1$ and $b_n=1$ and initial state the post collision state from the previous step,  corresponding to biasing the trajectory towards the outcome where the ancilla is found in the excited or the ground state respectively. Note that since we are interested in a single-step bias $G_n=\hat{1}$.
    \item Calculate the sensitivity of the corresponding measurement trajectories described above and choose the one with the highest value.
    \item Repeat until the desired number of collisions, $N$ has been achieved. As a result, the desired optimal trajectory $\boldsymbol{m}^{\rm max}$ is obtained.
    \item After obtaining the optimal trajectory construct the associated physical map and  compute the FI of the biased dynamics analytically or numerically by sampling sufficient number of quantum trajectories.
\end{enumerate}

Although this scheme does not yield as significant enhancement as the "global" optimization approach, as shown in Fig. \ref{fig:fig6}(c), it offers two key advantages. First, in contrast to the ``global" approach, the procedure does not need to be entirely repeated from the beginning when changing the total number of desired collisions, $N$. Additionally, it eliminates the need to compute the Fisher information (FI) of the initial physical map—a task that, while feasible, can be challenging for large system sizes or long evolution times. Finally, although here we focus on an approach based on one-step bias, generilization to a $n$-step bias is straightforward. 

\emph{Conclusions} \textendash{} In this work we have presented a systematic way to bias the quantum trajectories of monitored open quantum systems for enhanced metrology. A key element of our approach is the interpretation of the FI as an ensemble average of sensitivities of individual trajectories and engineer the interaction of the system with the environment such that trajectories with high sensitivity are more likely to occur. Focusing on two optimization algorithms, the ``global" and ``local" approach, we have shown that a great enhancement of the overall FI and thus precision can be achieved even for the simple case of a single two level atom used as a toy model. 

Our strategy can be straightforwardly implemented in current state-of-the-art quantum hardware platforms~\cite{Schindler2013, Arute2019, Evered2023,aquila} and can be adjusted in accordance to the requirements and the goals of the experiments of interest. Besides its promise to practical advancement in several metrological tasks, including protocols relying on continuous measurement, our work further raises some interesting theoretical questions for future work. 

First of all, it will be interested to investigate the effect on the FI of choosing different biasing functions, $\mathcal{B}$, which for example might permit bias of arbitrary $n$-point time correlations among measurements of the ancillas within the quantum trajectories. Furthermore, another promising perspective is to apply the scheme described in this work to complex quantum many-body systems and examine whether a precision beyond the shot noise limit can be more easily achieved. Finally, it is important to study the complexity of the required constructed physical map and further compare with alternative adaptive strategies~\cite{PhysRevA.77.052320,Pang2017,PhysRevX.7.041009,Sekatski_2017} that may result in the generation of alternative improved biasing optimization algorithms.

\emph{Acknowledgements } \textendash{} The author would like to thank D. Yang, D. Cillufo, M.B. Plenio and H. T{\"u}reci for fruitful discussions. Part of the simulations presented in this article were performed on computational resources managed and supported by Princeton Research Computing, a consortium of groups including the Princeton Institute for Computational Science and Engineering (PICSciE) and Research Computing at Princeton University.

\newpage
\onecolumngrid
\appendix
\section{Tilted Kraus operators in the small collision-time limit }
\label{app:small_collision_limit}
In this section we derive the expression of the tilted physical map described in Eqs. (5) and (6) of the main text in the limit where the time interval of a single collision $\Delta t \rightarrow 0$ corresponding to the continuous-time limit of our protocol. It is straightforward to show that in such regime, where we keep terms up to first order in $\Delta t$, the original Kraus operators take the simplified form

\begin{align}
&K_0= \langle \downarrow |\hat{U} |\downarrow \rangle=\hat{1}-i\omega \Delta t \hat{\sigma}_x-\frac{\Gamma}{2} \Delta t \hat{\sigma}_+ \hat{\sigma}_- \label{eq:Kraus0} \\
&K_1= \sqrt{\Gamma \Delta t} \hat{\sigma}_-, \label{eq:Kraus1}
\end{align}
which give rise to the Lindblad master equation (LME)

\begin{equation}
\frac{d \rho }{dt}=\mathcal{L}[\rho]=-i \omega[\hat{\sigma}_x, \rho]+ \Gamma \left(\hat{\sigma}_-\rho \hat{\sigma}_+ +\frac{1}{2} \{\hat{\sigma}_+ \hat{\sigma}_- ,\rho\} \right).
\end{equation}
Consequently the recursive relations in Eqs. (5) and (6) can be simplifies and arrive to the following lemma

\noindent \textbf{Lemma.} \textit{In the regime where $N \Gamma \Delta t \ll 1$ and  $\displaystyle \Gamma \Delta t \sum_{n=1}^i e^{-s_n} \ll 1$ for any value of $i$ and number of total collisions $N$  we have 
\begin{equation}
\label{eq:G_small}
G_{n-1}^2=\hat{1}- (N-n) \Gamma \Delta t \hat{\sigma}_+ \hat{\sigma}_- +\sum_{i=n+1}^N e^{-s_i} \Gamma \Delta t \hat{\sigma}_+ \hat{\sigma}_-.  
\end{equation}}

\vspace{2mm}
\noindent \textbf{Proof.} We will prove the above lemma by using the method of proof by induction. Eq. (\ref{eq:G_small}) is trivially true for $n=N$ since by construction $G_N=\hat{1}$. Let us assume now that it is also true for $n=k$. Then

\begin{align}
G_{k-1}^2&=K_0^{\dagger}G_k^2 K_0+e^{-s_k} K_1^{\dagger} G_k^2 K_1 \nonumber\\ 
&=K_0^{\dagger}\left[\hat{1}- (N-k) \Gamma \Delta t \hat{\sigma}_+ \hat{\sigma}_- +\sum_{i=k+1}^N e^{-s_i} \Gamma \Delta t \hat{\sigma}_+ \hat{\sigma}_-\right]K_0 \nonumber \quad \text{\small (by induction hypothesis)}\\
&+e^{-s_k}K_1^{\dagger}\left[\hat{1}- (N-k) \Gamma \Delta t \hat{\sigma}_+ \hat{\sigma}_- +\sum_{i=k+1}^N e^{-s_i} \Gamma \Delta t \hat{\sigma}_+ \hat{\sigma}_-\right]K_1 \nonumber \\
&=\hat{1}-2\frac{\Gamma}{2} \Gamma \Delta t \hat{\sigma}_+ \hat{\sigma}_- -(N-k) \Gamma \Delta t \hat{\sigma}_+ \hat{\sigma}_-+\sum_{i=k+1}^N e^{-s_i} \Gamma \Delta t \hat{\sigma}_+ \hat{\sigma}_- \nonumber \\
&+e^{-s_k} \Gamma \Delta t \hat{\sigma}_+ \hat{\sigma}_- \nonumber \quad \text{\small (by algebra using Eqs (\ref{eq:Kraus0}) and (\ref{eq:Kraus1})  and keeping terms up to first-order in $\Delta t$)} \\
&=\hat{1}- (N-(k-1)) \Gamma \Delta t \hat{\sigma}_+ \hat{\sigma}_- +\sum_{i=k}^N e^{-s_i} \Gamma \Delta t \hat{\sigma}_+ \hat{\sigma}_- \nonumber
\end{align}
Consequently by induction Eq. (\ref{eq:G_small}) is true for any $1\leq n \leq N$. As a result we can also write 

\begin{equation}
\label{eq:eq_ap8.4}
G_{n}^2=\hat{1}-A_n \Gamma \Delta t \hat{\sigma}_+ \hat{\sigma}_- \approx e^{-A_n \Gamma \Delta t \hat{\sigma}_+ \hat{\sigma}_-},
\end{equation}
where $A_n=(N-n)-\sum_{i=n+1}^N e^{-s_i}$.  Therefore after some algebra we arrive at the simplified expression of the tilted Kraus operators
\begin{align}
&\tilde{K}_0^{(n)}=\hat{1}-i\Omega \Delta t \hat{\sigma}_x-e^{-s_n}\frac{\Gamma}{2} \Delta t \hat{\sigma}_+ \hat{\sigma}_- \\
&\tilde{K}_1^{(n)}=\sqrt{e^{-s_n} \Gamma \Delta t}  \hat{\sigma}_+ \hat{\sigma}_-.
\end{align}
As a result, in the continuous monitoring limit of $\Delta t \rightarrow 0$ the new Kraus operators have exactly the same form as the original ones but with a transformed dissipation rate $\Gamma \rightarrow \Gamma e^{-s_n}$ which depends on the collision or equivalently on time. 

\bibliography{References}
\end{document}


\title{Supplemental Material-Criticality-enhanced Electromagnetic Field Sensor with Single Trapped Ions}

\author{Theodoros Ilias}
\author{Dayou Yang}
\author{Susana F. Huelga} 
\author{Martin B. Plenio}

\section{Realization of the phonon detector module }
\label{app:phonon_detector_module}

In this section we discuss in detail the implementation of the phonon detector module with the help of the ancilla-ion, which has an internal ladder-type three-level structure as shown in Fig.~\ref{fig:fig1}. Neglecting the system-ion, the Hamiltonian for the internal transition and the external motion (i.e., the phonon mode) of the ancilla reads ($\hbar=1$)
\begin{equation}
\label{eq:eq_ap_1}
\hat{H}_{\rm {a}}^{(I)}(t)=\omega_{\rm ph} \hat{c}^{\dagger}\hat{c}+\sum_{i=g,e,d} \omega_i | i\rangle \langle i |+ \Omega_w (|e\rangle \langle g|+\text{h.c.} ) \left[ e^{i (k_w \hat{x}-\omega_w t-\phi_w)} +\text{h.c.} \right]+\Omega_s (|d\rangle \langle e|+\text{h.c.} ) \left[ e^{i (k_s \hat{x}-\omega_s t-\phi_s)} +\text{h.c.} \right],
\end{equation}
where the weak (strong) laser beam is characterised by the Rabi frequency $\Omega_{w (s)}$, frequency $\omega_{w (s)}$, wavevector $k_{w (s)}$ and phase $\phi_{w (s)}$. Here $\hat{x}=x_0 (\hat{c}+\hat{c}^{\dagger})$ is the ancilla-ion position operator, with $x_0=1/\sqrt{2 \omega_{\rm ph}m_{\rm{a} }}$ and $m_{\rm {a} }$ the mass of the ion. We assume that the Lamb-Dicke parameter of the two lasers $\eta_{w (s)} :=k_{w (s)} x_0$ are the same and denote them as $\eta_{\rm LD}^{(2)}$. Moving to a frame rotating with respect to $\hat{H}_0=\omega_{\rm ph} \hat{c}^{\dagger} \hat{c}+(\omega_e-\omega_g) |e \rangle \langle e|+(\omega_d-\omega_g) |d \rangle \langle d|$, the Hamiltonian Eq.~\eqref{eq:eq_ap_1} takes the form
\begin{align}
\hat{H}_{\rm {a}}(t) &= \Omega_w (|e\rangle \langle g| e^{i(\omega_e-\omega_g)t}+\text{h.c.} ) \left[ e^{i (\eta_{\rm LD}^{(2)} (\hat{c}e^{-i \omega_{\rm ph}t}+\hat{c}^{\dagger}e^{i \omega_{\rm ph}t})-\omega_w t-\phi_w)} +\text{h.c.} \right] \\ 
&+\Omega_s (|d\rangle \langle e| e^{i(\omega_d-\omega_e)t}+\text{h.c.} ) \left[ e^{i (\eta_{\rm LD}^{(2)} (\hat{c}e^{-i \omega_{\rm ph}t}+\hat{c}^{\dagger}e^{i \omega_{\rm ph}t})-\omega_s t-\phi_s)} +\text{h.c.} \right]. \nonumber
\end{align}
In the Lamb-Dicke regime $\eta \sqrt{\langle (\hat{c}+\hat{c}^{\dagger})^2 \rangle} \ll 1$ we have 
\begin{equation}
e^{\eta_{\rm LD}^{(2)} (\hat{c}e^{-i \omega_{\rm ph}t}+\hat{c}^{\dagger}e^{i \omega_{\rm ph}t})} \approx \mathcal{I}+ \eta_{\rm LD}^{(2)}(\hat{c}e^{-i \omega_{\rm ph}t}+\hat{c}^{\dagger}e^{i \omega_{\rm ph}t}).
\end{equation}
We can tune the two laser frequencies such that the weak laser drives resonantly the red phonon sideband of the $\ket{g}\to\ket{e}$ transition, $(\omega_e-\omega_g)-\omega_w=\omega_{\rm ph}$; while the strong laser drives resonantly the carrier transition $\ket{g}\to\ket{d}$, $(\omega_d-\omega_e) -\omega_s=0$.  After choosing $\phi_s=\phi_w=0$, we perform an optical rotating-wave approximation (RWA) by neglecting terms rotating at frequency $\sim (\omega_e-\omega_g)+\omega_w$, $ \sim (\omega_d-\omega_e)+\omega_s$ and a vibrational RWA where we neglect terms rotating at frequency $\sim \omega_{\rm ph}$. As a result, we obtain
\begin{equation}
\label{eq:Hamiltonian_a-i}
\hat{H}_{\rm {a}}=\Omega_{s} (| d\rangle \langle e|+|e\rangle \langle d |)+\eta_{\rm LD}^{(2)}\Omega_{w}(|e\rangle \langle g| \hat{c}+|g\rangle \langle e| \hat{c}^{\dag}).
\end{equation}
for the Hamiltonian of the trapped ancilla-ion. 

The quantum state of the ancilla-ion $\rho_{\rm {a}}$ (including both the internal transition and  the external motion) evolves according to the LME
\begin{equation}
\label{eq:LME_a-i}
\frac{ d \rho_{\rm {a}} } {dt}= -i [\hat{H}_{\rm {a}},\rho_{\rm {a}}]+ \Gamma_s \left( |e \rangle \langle d | \rho_{\rm {a}} |d \rangle \langle e| -\frac{1}{2} \{ |d\rangle \langle d|,  \rho_{\rm {a}}\}  \right)+\Gamma_w \left( |g \rangle \langle e | \rho_{\rm {a}} |e \rangle \langle g| -\frac{1}{2} \{ |e\rangle \langle e |,  \rho_{\rm {a}} \}  \right).
\end{equation}
To visualize such an evolution, in Fig.~\ref{fig:supp_fig1}(a) we plot the phonon mode occupation in the appropriate parameter regime see below, which demonstrates an exponential decay with time at an effective dissipation rate $\kappa$. To acquire an understanding, let us relate the effective decay to the microscopic parameters $\Omega_s$, $\Gamma_s$, $\Gamma_w$. We can divide one full cooling cycle $|g\rangle |n \rangle \rightarrow |g\rangle |n-1 \rangle$, where $|n \rangle$ denotes the phonon occupation number, into different elementary transitions. First, for small enough $\Gamma_w$, the event $|g\rangle |n \rangle \rightarrow |e \rangle | n-1 \rangle $ defines the transition from the dark to the bright state of the ancilla ion at a characteristic timescale $T_1 \approx \Omega_s^2/(\Omega_w^2 \Gamma_s)$~\cite{RevModPhys.70.101}.  The inverse, undesired transition $|e\rangle |n-1 \rangle \rightarrow |g \rangle | n \rangle $, with a characteristic timescale $T_2 \approx (\Gamma_s^2+\Omega_s^2) T_1/ \Gamma_s^2$, restores the initial state of the ancilla-ion but without any phonon annihilation. As a result, to realize an effective dissipation of the phonon mode, we require that the spontaneous emission $|e\rangle |n-1 \rangle \rightarrow |g\rangle |n-1 \rangle$ happens at a timescale $ 1/\Gamma_w \equiv T_3  \ll T_2$.  The time of a complete cooling cycle is given by $1/\kappa \equiv T_c\approx T_1+T_3 \approx T_1$ as we are in the regime where $\Omega_s \gg \Omega_w$. Such analytical understanding is confirmed by the numerical simulation shown in Fig.~\ref{fig:supp_fig1}(b), where we plot the dependence of  the effective dissipation rate $\kappa$ of the phonons with respect to the different parameters of our model.

\begin{figure}[t]
\centering{} \includegraphics[width=0.85\textwidth]{Supp_fig1.pdf} 
\caption{ Induced dissipation of the bosonic mode via a three level ancilla-ion. a) Comparison of the evolution of the phonon's excitation between an induced dissipation via an ancilla-ion with $\Gamma_w=40$, $\Omega_w=14.3 \Gamma_w$, $\Gamma_s=80 \Gamma_w$, $\Omega_s=2 \Gamma_s$ and $\eta_{\rm LD}^{(2)}=0.07$ (blue dotted line) and a direct dissipation with $\kappa=1.5\times10^{-3} \Gamma_w$ (orange solid line).  b) Dependence of the induced dissipation rate, $\kappa$, with respect to different adjustable parameters in our system. The plots indicate that $\kappa \sim \Omega_s^2/(\Omega_w^2 \Gamma_s)$ as expected from our physical understanding.}
\label{fig:supp_fig1} 
\end{figure}

\section{Trapped-ion implementation of the open Rabi model }
\label{app:real_rabi_model}

Here, we discuss the implementation of the open Rabi model as presented schematically in Fig.~\ref{fig:fig1}. The system-ion is co-trapped with the ancilla-ion analyzed in Appendix \ref{app:phonon_detector_module}, and is driven by two travelling-wave laser beams of the same Rabi frequency, $\Omega_0$, and Lamd-Dicke parameter, $\eta_{\rm LD}$. As a result the Hamiltonian of the total system reads

\begin{equation}
\hat{H}_{\rm tot}^{(I)}(t)=\hat{H}_{\rm {a}}^{(I)}(t)+\hat{H}_{\rm {s}}^{(I)}(t),
\end{equation}
where $\hat{H}_{\rm {s}}^{(I)}(t)=\frac{\omega_I}{2}\hat{\sigma}_z+\sum_j \frac{\Omega_0}{2} \hat{\sigma}_x \left( e^{i \eta_{\rm LD} (\hat{c}+\hat{c}^{\dagger})-\omega_j t -\phi_j }+\text{h.c.} \right)$ and $\hat{\sigma}_{z,x}$ are the Pauli matrices referring to the system-ion. In the rotating frame with respect to $\tilde{H}_0=\omega_{\rm ph} \hat{c}^{\dagger} \hat{c}+(\omega_e-\omega_g) |e \rangle \langle e|+(\omega_d-\omega_g) |d \rangle \langle d|+\frac{\omega_I}{2}\hat{\sigma}_z$, following similar approximations as before by choosing $\omega_1=\omega_I+\omega_{\rm ph}-\delta_{\rm b}$, $\omega_2=\omega_I-\omega_{\rm ph}-\delta_{\rm b}$,    $\phi_1=\phi_2=3 \pi /2$, $\omega_s=(\omega_d-\omega_g)$, $(\omega_e-\omega_g)-\omega_w=\omega_{\rm ph}+(\delta_{\rm b}-\delta_{\rm b})/2$ we arrive at 
\begin{align}
\hat{H}_{\rm tot}^{(I)}(t)&=\Omega_{s} (| d\rangle \langle e|+|e\rangle \langle d |)+\eta_{\rm LD}^{(2)}\Omega_{w}(|e\rangle \langle g| \hat{c}e^{i\frac{\delta_{ \rm b}-\delta_{\rm r}}{2}t}+|g\rangle \langle e| \hat{c}^{\dag}e^{-i\frac{\delta_{ \rm b}-\delta_{\rm r}}{2}t})\nonumber \\
&-\frac{\Omega_0 \eta_{\rm LD}}{2} ( \hat{\sigma}_+ e^{i\frac{\delta_{ \rm b}+ \delta_{ \rm r} }{2}t}+\text{h.c})(\hat{c}^{\dagger} e^{i\frac{\delta_{ \rm b}- \delta_{ \rm r} }{2}t}+\text{h.c}).
\label{eq:eq_ap_7}
\end{align}
In the frame rotating with $H^{'}_0=\frac{\delta_{\rm b}-\delta_{\rm r}}{2}\hat{c}^\dag\hat{c}+\frac{\delta_{\rm r}+\delta_{\rm b}}{4}\hat{\sigma}_z$, the evolution of the full model is governed by the LME
\begin{equation}
\label{eq:LME_a-i_s-i}
\frac{ d \rho } {dt}= -i [\hat{H}_{\rm {s}}+\hat{H}_{\rm {a}},]+ \Gamma_s \left( |e \rangle \langle d | \rho |d \rangle \langle e| -\frac{1}{2} \{ |d\rangle \langle d|,  \rho\}  \right)+\Gamma_w \left( |g \rangle \langle e | \rho |e \rangle \langle g| -\frac{1}{2} \{ |e\rangle \langle e |,  \rho \}  \right),
\end{equation}
which correctly implements the LME~({\ref{eq:LME_Rabi}}) of the open Rabi model as illustrated in Fig.~\ref{fig:supp_fig2}.

\begin{figure}[b!]
\centering{} \includegraphics[width=0.6\textwidth]{Supp_fig2.pdf} 
\caption{ Trapped-ion realisation of the open Rabi model. a) Dynamics of the phonon's excitation in the open Rabi model as implemented with the help of an ancilla-ion and evolved according to Eq. (\ref{eq:LME_a-i_s-i}) (blue-dotted line) and Eq. (\ref{eq:LME_Rabi}) (orange solid line).  Parameters: $\eta=50$, $\omega=0.6$; the rest kept the same as Fig.~\ref{fig:supp_fig1} .  b) Steady state occupation of the bosonic mode with respect to different system sizes, $\eta$, as calculated from Eq.(\ref{eq:LME_a-i_s-i})  (blue-dotted line) and Eq.(\ref{eq:LME_Rabi}) (orangle-dotted line). The excellent agreement between the two lines in both plots demonstrates the accurate implementation of the open Rabi model in our set-up.}
\label{fig:supp_fig2} 
\end{figure}

\section{Enhanced Continuous Phonon Detection}
\label{app: en_con_ph_det}
In this section we discuss the highly efficient continuous measurement of the phonon mode accomplished by the use of the ancilla-ion. Adopting the physical picture developed in Appendix \mbp{\ref{app:real_rabi_model}}, it is clear that once an event $|g\rangle |n \rangle \rightarrow |e \rangle | n-1 \rangle $ occurs, the strongly driven transition $|e\rangle \leftrightarrow | d\rangle$ of the ancilla-ion leads to the emission of photons at a rate $\Gamma_{\rm ph}=\Gamma_s \times P_{\rm ex} \approx  \frac{\Gamma_s}{2} \frac{(2 \Omega_s/\Gamma_s)^2}{1+(2 \Omega_s/\Gamma_s)^2}$, where $P_{\rm ex}$ is the probability to find the ancilla-ion in the $|d\rangle \langle d|$ state. The characteristic timescale for restoring the population, $|e\rangle \rightarrow |g\rangle$, is $T_2=1/\Gamma_w$. As a result, during a full cooling cycle $|g\rangle |n\rangle \rightarrow |g\rangle |n-1\rangle$ there are $N_{\rm em}=T_2 \times \Gamma_{\rm ph}\approx \Gamma_s/\Gamma_w$ photons emitted and $N_{\rm ph}=\epsilon N_{\rm em} $ photons detected which lead to the definition of the enhancement factor as in Eq. (\ref {eq:enhancement_factor}) of the main text. As a simple demonstration, let us again neglect the system-ion, and consider the enhanced measurement of a single vibrational phonon mode occupying an excited Fock state. The conditional evolution of the phonon mode is given by a stochastic master equation similarly to Eq.~(\ref{eq:sme}) of the main text. As shown in Fig.~\ref{fig:supp_fig3}, for any photon detector of efficiency, $\epsilon$, we can implement almost perfect detection of the phonon mode by appropriate engineer of the parameters for the internal transitions the ancilla.

\begin{figure}[t!]
\centering{} \includegraphics[width=0.5\textwidth]{Supp_fig3.pdf} 
\caption{Conditional evolution of the damped bosonic mode of the trap as implemented with the help of an ancilla-ion where each annihilated phonon is accompanied with the emission and detection of lots of photons. Upper panel: The photon-detection signal, $D(t,0)$, of a detector with efficiency $\epsilon=0.5$; Lower panel: the conditional evolution of the damped mode as exemplified by the bosonic mode occupation $\langle \hat{c}^{\dagger} \hat{c} \rangle$. The dynamics of a single trajectory associated with the corresponding detected signal is the same as the ideal case of perfect photon detection. Parameters: $\Gamma_w=40$, $\Omega_w=14.3\Gamma_w$, $\Omega_s=2\Gamma_s=160 \Gamma_w$ and $\eta_{\rm LD}^{(2)}=0.07$.}
\label{fig:supp_fig3} 
\end{figure}

\section{Quality of the Scaling Collapse}
The finite-size scaling analysis is a powerful numerical method to extract the relevant exponents of continuous phase transitions. Consider a general quantity $\mathcal{A}$ dependent on two parameters $h$ and $L$ according to
\begin{equation}
\mathcal{A}(h,L)=h^{a} f(h/L^{b}).
\label{eq:scaling}
\end{equation}
Depending on the nature of the model system, $\mathcal{A}$, $L$ and $h$ can refer to different quantities.  For example, $\mathcal{A}$ might refer to the magnetization of the Ising spin chain, with $h$ being the inverse of the transverse magnetic field and $L$  the length of the chain. In our case, $\mathcal{A}$ refers to the Fisher information $F_{\omega}(t, \eta)$, with $L$ and $h$ being the system size $\eta$ and the evolution time $t$ respectively.  From Eq.~(\ref{eq:scaling}), it is clear that if we plot $\mathcal{A} h^{-a}$ against $h/L^{b}$ for different values of $L$ and $h$, all the curves collapse onto a single. Therefore, we can determine the unknown exponents $a$ and $b$ by numerical fitting of the data to Eq.~\eqref{eq:scaling} and find the best scaling collapse. 

We can define appropriate measure of the quality of the data collapse to remove subjectiveness of the approach. If the scaling function $f(x)$ is known, we can define the measure~\cite{Somendra}
\begin{equation}
\label{eq:measure_data_collapse1}
\mathcal{M}_{kn}(a,b)=\sqrt{\frac{1}{N} \sum_{ij} \left( \frac{\mathcal{A}({L_i,h_j})h_j^{-a}-f(L_i^{-b}h_j)}{f(L_i^{-b}h_j)}\right)^2}
\end{equation}
which is minimized by the optimal $a$ and $b$, with $N$ the total number of data points. Notice that in Eq.~(\ref{eq:measure_data_collapse1}) the division with respect to the scaling function $f(x)$ is essential---otherwise $\mathcal{M}_{kn}$ can be minimized for small values of the numerator which not necessarily indicates a good collapse.

In the general case the scaling function $f(x)$ is not known. We can interpolate $f(x)$ via any set of data points corresponding to a specific $L$. Denoting the different sets via the subscript $p$, we can measure the quality of the data collapse by comparing the data points of pairs of sets $p_{1,2}$ in their overlapping regions and summing up all the contribution,
\begin{equation}
\label{eq:mesure_data_collapse2}
\mathcal{M}(a,b)=\sqrt{ \frac{1}{N}\sum_{p} \sum_{i\neq p} \sum_{j, \rm{ov}} \left(\frac{\mathcal{A}({L_i,h_j})h_j^{-a}-\mathcal{E}_p(L_i^{-b}h_j)}{\mathcal{E}_p(L_i^{-b}h_j)}\right)^2},
\end{equation}
where $\mathcal{E}_p(x)$ is the interpolation function based on the basis set-$p$ and $N$ is the total number of points in all overlapping regions. Here, the innermost index $j$ runs over the overlapping region of the set-$p$ and another set-$i$. It is clear that   $\mathcal{M} \geq 0$ with the zero lower bound achieved only in case of perfect data collapse. As a result, it can be used via the variational principle for an automatic and objective extrapolation of the relevant critical exponents.

For our case, in order to examine the quality of the finite-size scaling analysis for different decoherence rates, as shown in Fig.~\ref{fig:fig3}(b) in the main text, we introduce the quality factor
\begin{equation}
Q=\frac{\mathcal{M}^F_{\rm id}} {\mathcal{M}^F},
\end{equation}
where $ \displaystyle \mathcal{M}^F=\mathcal{M}(2,1)=\sqrt{ \frac{1}{N}\sum_{p} \sum_{i\neq p} \sum_{j, \rm{ov}} \left(\frac{F (t_j \eta_i) t_j^2-\mathcal{E}_p(\eta_i t_j)}{\mathcal{E}_p(\eta_i t_j)}\right)^2}$ is the measure of the data collapse, and $\mathcal{M}^F_{\rm id}$ refers to the measure of the ideal noiseless case.